\documentclass[a4paper,11pt]{article}
\usepackage{pos}

\newcommand{\Hii}{H~{\sc ii}}

\newcommand{\kms}{km\,s$^{-1}$}
\newcommand{\pcmc}{cm$^{-3}$}
\newcommand{\pcms}{cm$^{-2}$}

\title{Deuterated molecules in regions of high-mass star formation}

\author*[a]{Igor I. Zinchenko}
\author[a,b]{Andrey G. Pazukhin}
\author[a]{Elena A. Trofimova}
\author[a]{Peter M. Zemlyanukha}
\author[c,d]{Christian Henkel}
\author[e]{Magnus Thomasson}

\affiliation[a]{Institute of Applied Physics of the Russian Academy of Sciences,\\
  46 Ul'yanov Street, Nizhny Novgorod, 603950, Russia}

\affiliation[b]{Faculty of Radiophysics, Lobachevsky State University of Nizhny Novgorod,\\
23 Gagarin Ave, 603950, Niznij Novgorod, Russia}

\affiliation[c]{Max-Planck-Institut f\"ur Radioastronomie,\\
  Auf-dem-H\"ugel 69, 53121, Bonn, Germany}

\affiliation[d]{Astron. Dept., Faculty of Science, King Abdulaziz University,\\
PO Box 80203, Jeddah, 21589, Saudi Arabia}

\affiliation[e]{Department of Space, Earth and Environment, Chalmers University of Technology,\\
  Onsala Space Observatory, SE-43992 Onsala, Sweden}

\emailAdd{zin@iapras.ru}
\emailAdd{pazukhinandrey@bk.ru}
\emailAdd{tea@iapras.ru}
\emailAdd{petez@iapras.ru}
\emailAdd{chenkel@mpifr-bonn.mpg.de}
\emailAdd{magnus.thomasson@chalmers.se}

\abstract{We present the results of our studies of deuterated molecules (DCN, DNC, DCO$^+$, N$_2$D$^+$ and NH$_2$D) in regions of high-mass star formation, which include a survey of such regions with the 20-m Onsala radio telescope and mapping of several objects in various lines with the 30-m IRAM and 100-m MPIfR radio telescopes. The deuteration degree reaches $\sim$10$^{-2}$ in these objects. We discuss its dependencies on the gas temperature and velocity dispersion, as well as spatial distributions of deuterated molecules. We show that the H$^{13}$CN/HN$^{13}$C intensity ratio may be a good indicator of the gas kinetic temperature and estimate densities of the investigated objects.}

\FullConference{%
 The Multifaceted Universe: Theory and Observations --  2022 (MUTO2022)\\
  23--27 May 2022\\
  SAO RAS, Nizhny Arkhyz, Russia\\
}

\begin{document}
\maketitle

\section{Introduction}
The effect of deuteration of molecular gas in interstellar clouds (i.e., an increase of the relative abundance of deuterated molecules) has been investigated for a long time already. It is related to the exothermicity of the reactions of replacing a proton with deuterium in molecules, which underlie the chains of the chemical reactions leading to the formation of most other molecules (e.g., \cite{Roueff07}). The most important among them is the following reaction:
\begin{equation}
\mathrm{H_3^+ + HD} \rightleftharpoons \mathrm{H_2D^+ + H_2} + 232\,\mathrm{K} \, .
\end{equation}
In addition, at low temperatures, freezing of molecules which destroy H$_2$D$^+$, in particular CO, on dust grains is important, as well as the decreased ionization degree which reduces the recombination rate of H$_2$D$^+$.

Deuteration has been studied mostly in cold interstellar clouds, where low mass stars form. In warmer high-mass star-forming (HMSF) regions it is expected to be much weaker due to the strong temperature dependence of the factors mentioned above. Nevertheless, it has been observed in such regions, too \cite[e.g.][]{Pillai07, Fontani11,  Miettinen11}. Several years ago a survey of 59 such regions in the lines of some deuterated molecules was performed \cite{Gerner15}. It was found that the deuteration degree depends on the evolutionary stage of the objects and correlates with the luminosity of the central source. Detailed chemical models are available, which describe the expected abundance of deuterated molecules \cite[e.g.][]{Albertsson13}.

Studies of this effect are based on observations of the low-excitation rotational transitions of deuterated molecules and their isotopologues. However, the lowest rotational transitions of DCN, DNC, DCO$^+$, N$_2$D$^+$ fall into a wavelength range near 4~mm, which is rather poorly covered in radio astronomy due to the presence of a strong absorption band of atmospheric oxygen near 60~GHz. At the same time the observations of the lowest rotational transitions are important for more accurate determination of the total column density. Nowadays this band is available at the 20-m Onsala radio telescope (Sweden) and 30-m IRAM radio telescope on Pico Veleta (Spain). In recent years we performed a survey of several tens of HMSF regions at 3--4~mm with the 20-m telescope and a more detailed study of several regions with the 30-m telescope at 3--4~mm and 2~mm. In addition, several sources have been mapped with the 100-m telescope at Effelsberg (Germany) in the ammonia lines to get information on the gas kinetic temperature. The results of the survey have been partly published \cite{Trofimova20}, and other publications are in preparation (Pazukhin et al., Trofimova et al.). Here we summarize the main results of these studies.

\section{Observational data}
As mentioned above, we analyze the data obtained with three radio telescopes: the 20-m Onsala radio telescope, the 30-m IRAM radio telescope and the 100-m radio telescope in Effelsberg. Details of the Onsala observations are presented in \cite{Trofimova20}. Briefly, about 60 sources in total were observed in 2017 and 2018 at Onsala. They were selected by the presence of certain signs of high mass star formation such as H$_2$O masers, powerful IR sources, and \Hii\ zones. Two receiver settings were used, in which the following frequency bands were covered: (1) 71.94--74.44 GHz, (2) 75.45--77.95 GHz, (3) 83.94--86.44 GHz, and (4) 87.45--89.95 GHz. The bands (1, 3) or (2, 4) were observed simultaneously. A spectral resolution of 76 kHz (which corresponds to $\sim$0.3~\kms) was used. The half power beam width (HPBW) was between $\approx$50$^{\prime\prime}$ and $\approx$40$^{\prime\prime}$ at these frequencies. The line list includes the $J=1-0$ transitions of DCN, DNC, DCO$^+$, N$_2$D$^+$ and the ortho-NH$_2$D $1_{11}-1_{01}$ line. In addition, several other important lines were observed, in particular the CH$_3$CCH line series, which can be used as an indicator of the gas kinetic temperature \cite{Bergin94, Malafeev05}. 

The observations with the 30-m IRAM radio telescope and the 100-m radio telescope in Effelsberg are described in Pazukhin et al. (in preparation). The observations at the 30-m telescope were performed in 2019 in the 3--4~mm and 2~mm bands simultaneously. Six sources were mapped in the On-The-Fly total power mode. Both $J=1-0$ and $J=2-1$ transitions of DCN, DNC, DCO$^+$ and N$_2$D$^+$ were observed. Other isotopologues of these molecules were observed, too. Several other important lines were observed including two CH$_3$CCH line series (at 3~mm and 2~mm). The HPBW was from $\sim30^{\prime\prime}$ at 4~mm to $\sim16^{\prime\prime}$ at 2~mm.

Making use of the 100\,m telescope near Effelsberg (Germany), we performed observations of the H$_2$O 22\,GHz maser transition and the ammonia (1,1), (2,2) and (3,3) inversion lines during 2019, December 9. The measurements were carried out in a continuous mapping mode (On-The-Fly) using a K-band secondary focus receiver with a dual bandwidth of 300\,MHz, including the above mentioned H$_2$O in one and the NH$_3$ transitions in the other band. We obtained $5^\prime \times 5^\prime$ maps of the sources. The HPBW of the observations was $\sim40^{\prime\prime}$.

\section{Main results}
\subsection{The survey results} \label{sec:survey}
For the Onsala survey, the detection statistics are as follows: In 16 out of 50 sources we detected DCO$^+$ in band (1). DCN molecules were visible in 17 of these sources. The frequency band (2) contains a total of 47 sources. In 15 of them DNC has been detected, while N$_2$D$^+$ was only seen in two. Towards 15 out of 50 observed sources we detected NH$_2$D in the frequency band (3).

The column densities of the molecules were estimated with a non-LTE model using the publicly available RADEX program \cite{Radex}. When the data for deuterated molecules were not available in the LAMDA database \cite{LAMDA} used by RADEX, the data from other isotopologues were used. When possible, the gas kinetic temperature was estimated from the CH$_3$CCH line series by the rotation diagram method. Otherwise it was taken from publications. The gas volume density was assumed to be $10^5$~\pcmc. This value is close to the so-called critical density for these molecules. In this case the estimate of the column density is close to the minimum value \cite{Trofimova20}. 

\begin{figure}[h]
\begin{minipage}[t]{0.325\linewidth}
\center{\includegraphics[width=\linewidth]{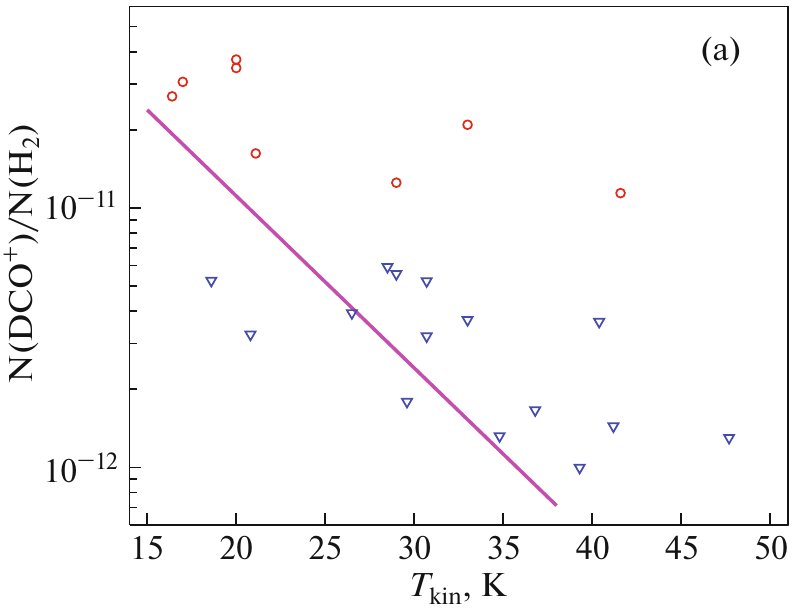}} 
\end{minipage}
\hfill
\begin{minipage}[t]{0.325\linewidth}
\center{\includegraphics[width=\linewidth]{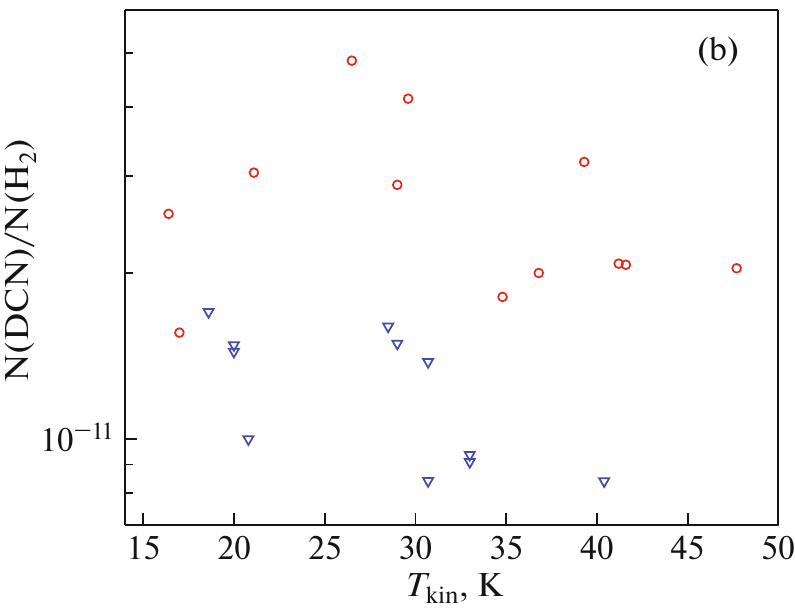}} 
\end{minipage}
\hfill
\begin{minipage}[t]{0.325\linewidth}
\center{\includegraphics[width=\linewidth]{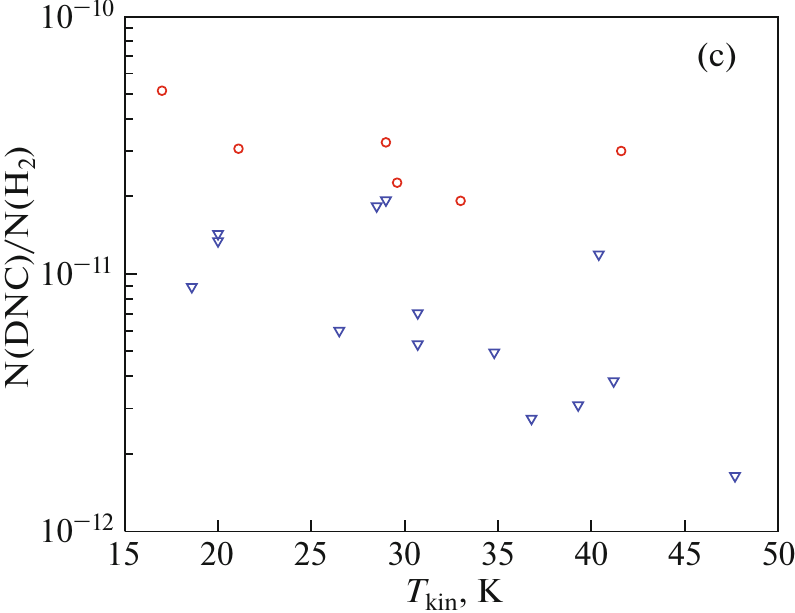}} 
\end{minipage}
\caption{The dependence of relative abundances on kinetic temperature following molecules: (a) DCO$^+$, (b) DCN, (c) DNC. The triangles represent upper limits for the corresponding molecules. The purple line is a linear regression fit also considering upper limits according to \cite{Cor86}. The plots are adopted from \cite{Trofimova20}.}
\label{ris:N/H-Tkin}
\end{figure}

The molecular hydrogen column densities were estimated from the C$^{18}$O data presented in \cite{Zin00} assuming a C$^{18}$O relative abundance of $1.7\times 10^{-7}$ \cite{Frer82}. Then the relative abundances of the deuterated molecules were calculated. The dependencies of these abundances on temperature and velocity dispersion were analyzed taking into account data represented by upper and lower limits \cite{Cor86}. The dependencies on temperature are presented in Fig.~\ref{ris:N/H-Tkin}. A statistically significant decrease of the DCO$^+$ abundance with increasing temperature was found, while the DCN abundance remains nearly constant. The trend for decreasing DNC abundance in Fig.~\ref{ris:N/H-Tkin}c is not statistically significant.

\begin{figure}[h]
\begin{minipage}[t]{0.325\linewidth}
\center{\includegraphics[width=\linewidth]{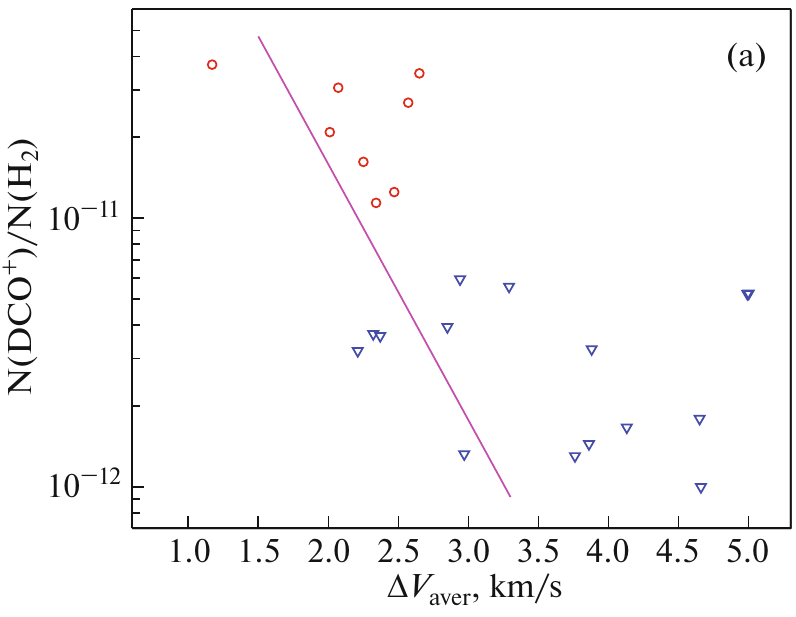}} 
\end{minipage}
\hfill
\begin{minipage}[t]{0.325\linewidth}
\center{\includegraphics[width=\linewidth]{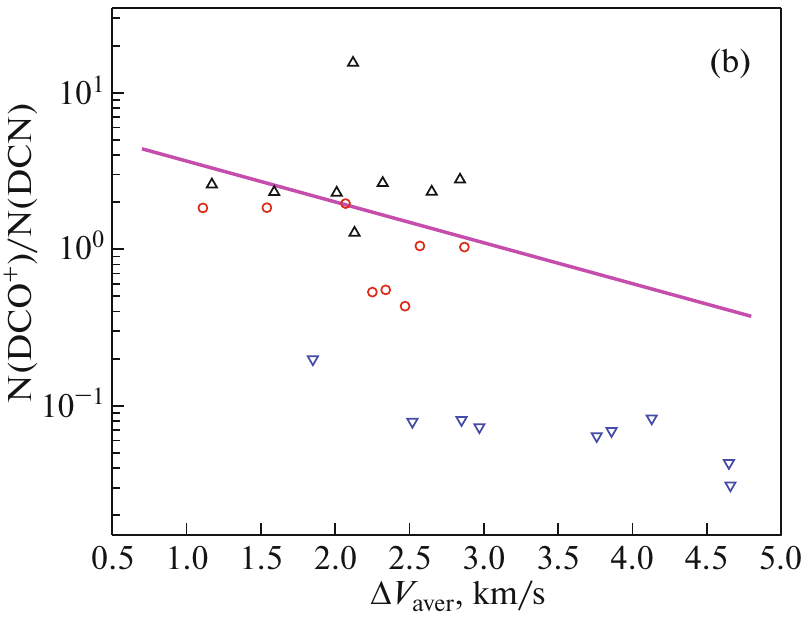}} 
\end{minipage}
\hfill
\begin{minipage}[t]{0.325\linewidth}
\center{\includegraphics[width=\linewidth]{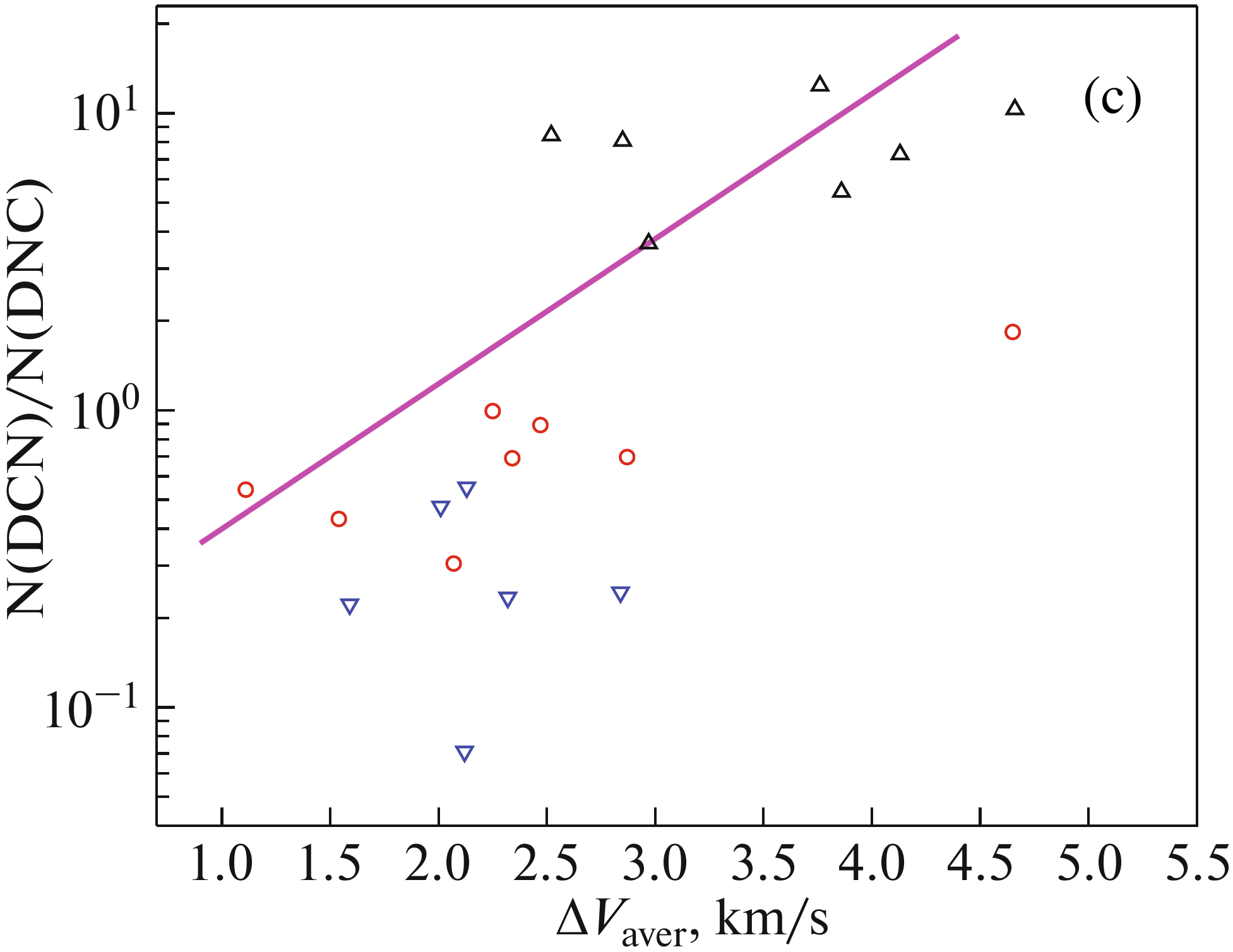}} 
\end{minipage}
\caption{The dependence of column density ratios on the average width of narrow lines in our sources: (a) N(DCO$^+$)/N(H$_2$), (b) N(DCO$^+$)/N(DCN), (c) N(DCN)/N(DNC). On all dependencies the upward triangles are lower limits, the downward triangles are upper limits. The straight purple lines represent linear regression fits accounting for both upper and lower limits according to \cite{Cor86}. The plots are adopted from \cite{Trofimova20}.}
\label{fig:dv-dep}
\end{figure}

We also considered dependencies of the molecular abundances and ratios of abundances on the width of narrow lines in the source, which serves as a measure of the velocity dispersion (Fig.~\ref{fig:dv-dep}). There is a noticeable decrease of the DCO$^+$ abundance with increasing line width. The relative abundances of some molecules also vary significantly. Similar trends (although non-conclusive) were presented in \cite{Gerner15}. It is worth noting that there is a correlation between the line width (which is non-thermal in these sources) and the gas kinetic temperature \cite{Trofimova20}. It is natural because the temperature is determined by the luminosity of the central source mainly. The sources with a higher luminosity have a stronger effect on the surrounding medium increasing gas turbulence.

To estimate the deuteration degree, observations of the hydrogenated isotopologues are needed. However, the optical depth in the HCN and HCO$^+$ lines is very high, which prevents a reliable determination of their column densities. For this reason we used the H$^{13}$CN and HC$^{18}$O$^+$ $J=1-0$ lines, which are present in our data. However the HC$^{18}$O$^+$ line was detected only in a few sources and only in two sources both DCO$^+$ and HC$^{18}$O$^+$ lines were detected. For them the DCO$^+$/HCO$^+$ abundance ratio is $\sim$8.5$\times 10^{-3}$. The DCN/HCN ratio could be determined for a much larger set of sources. It is $\sim$10$^{-2}$ for the sources detected in the DCN line and remains nearly constant in the temperature range $15-50$~K. The results were derived using isotopologue ratios $^{16}$O/$^{18}$O$\sim$500 and $^{12}$C/$^{13}$C=65 \cite{Trofimova20}. These results are consistent with conclusions that the DCN molecules can form at higher temperatures than DCO$^+$ and DNC \cite{Albertsson13}. The NH$_2$D/NH$_3$ ratio also remains practically constant ($\sim$10$^{-2}$) in this temperature range (Trofimova et al., in preparation), which contradicts the chemical model for ammonia deuteration presented in \cite{Roueff05}, predicting 
the NH$_2$D/NH$_3$ ratio $\sim$0.1 at 10--20~K and its drop to $\sim 2\times 10^{-3}$ at 50~K.

\subsection{Maps of the sources}

\begin{figure}[h]
\begin{minipage}[t]{0.325\linewidth}
\center{\includegraphics[width=\linewidth]{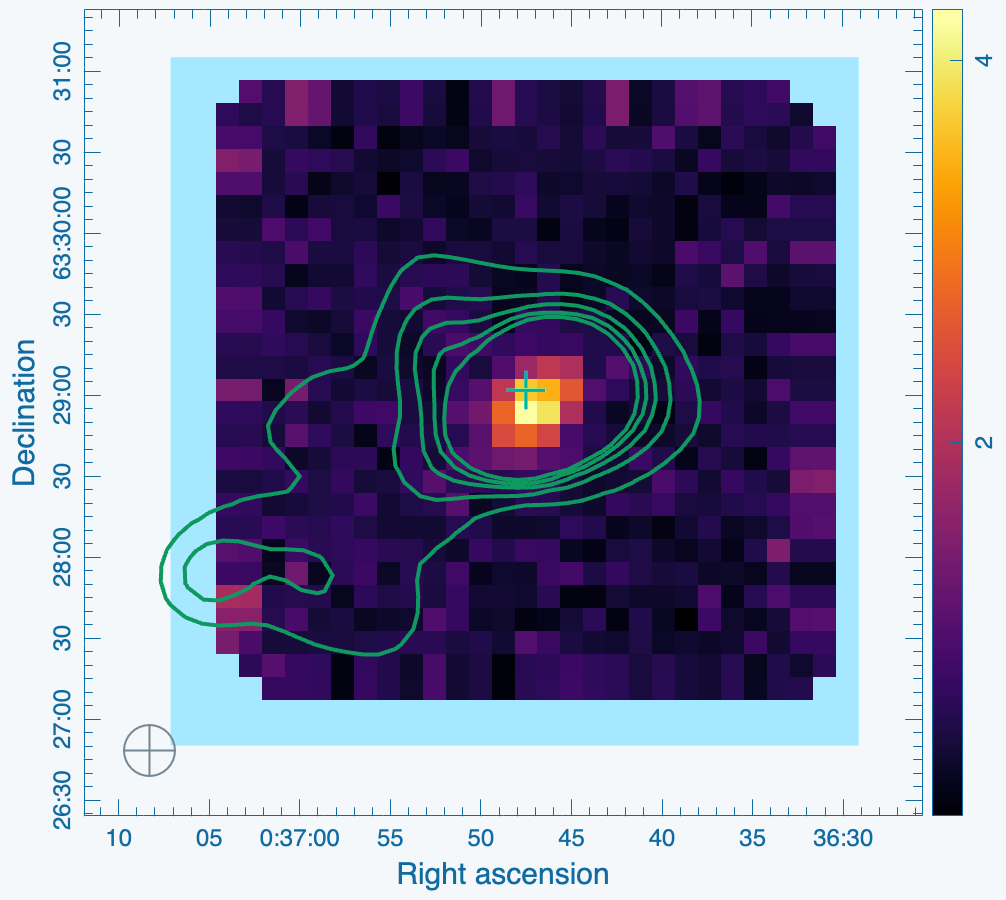}} 
\end{minipage}
\hfill
\begin{minipage}[t]{0.325\linewidth}
\center{\includegraphics[width=\linewidth]{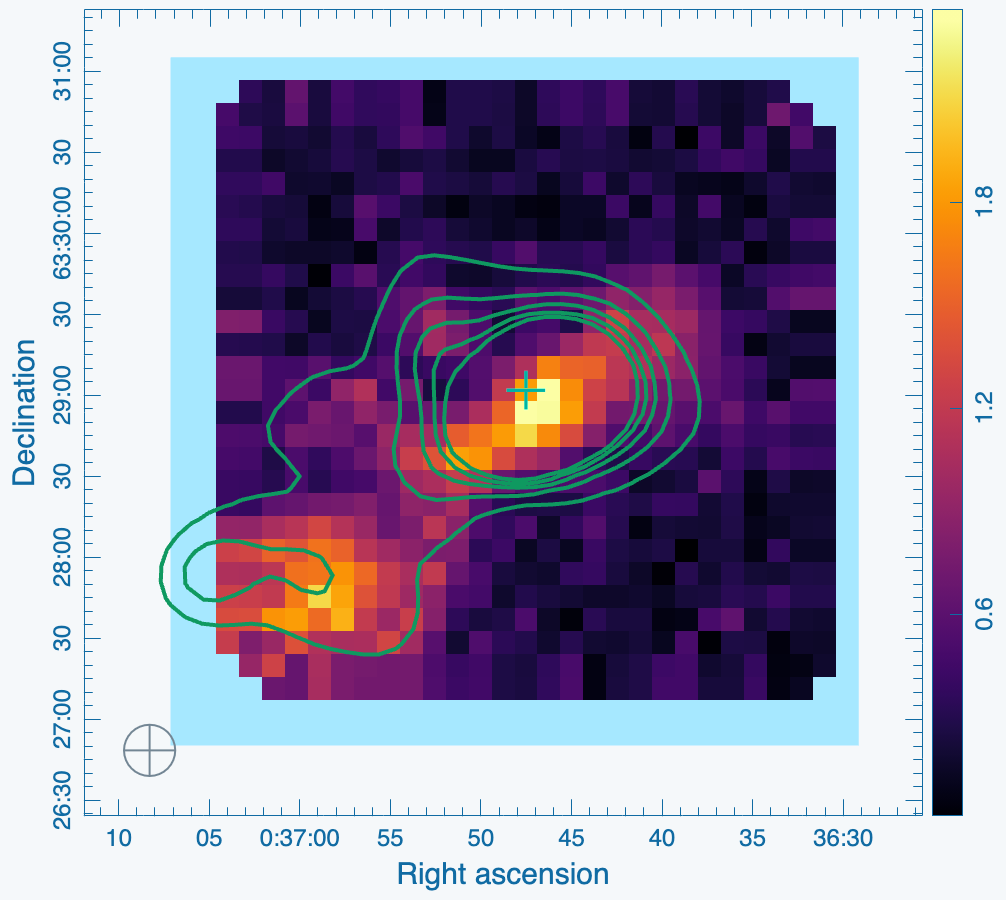}} 
\end{minipage}
\hfill
\begin{minipage}[t]{0.325\linewidth}
\center{\includegraphics[width=\linewidth]{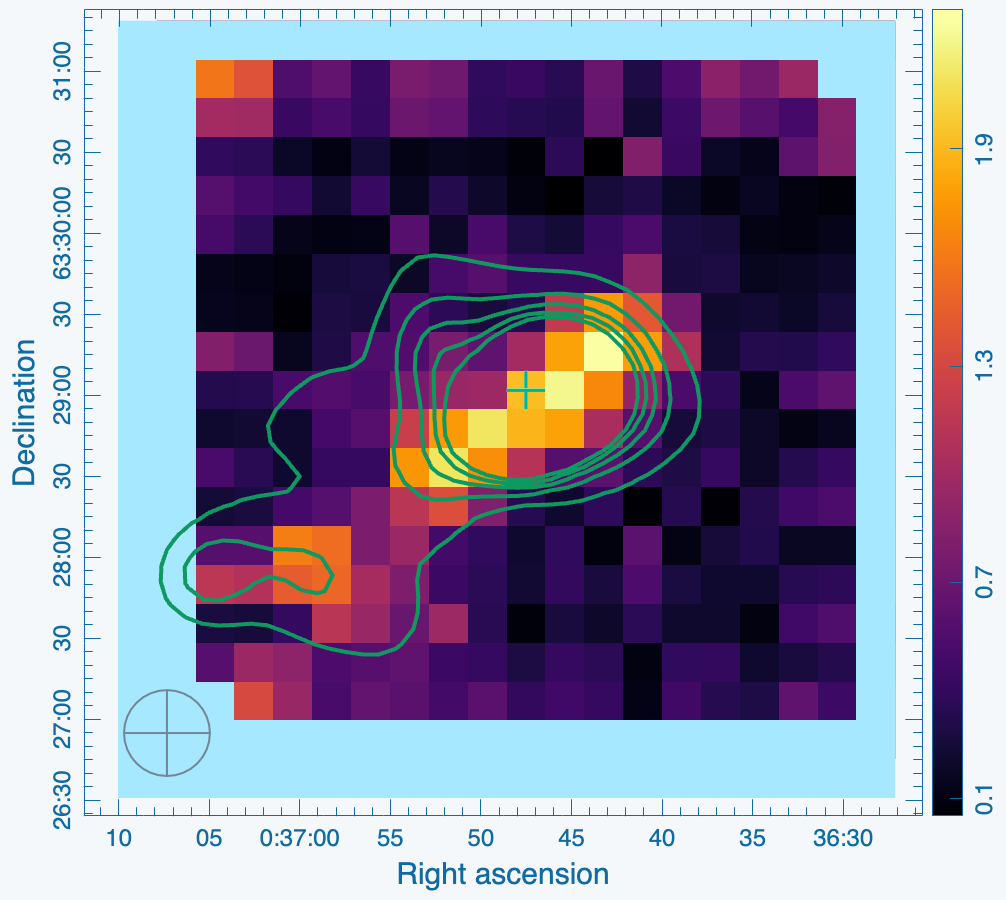}} 
\end{minipage}
\caption{Maps of L1287 in the DCN, DCO$^+$ $J=2-1$ and NH$_2$D $1_{11}-1_{01}$ lines (from left to right) obtained with the 30-m IRAM radio telescope. The contours represent the H$_2$ column density 
    obtained from the \textit{Herschel} data by the PPMAP algorithm \cite{Marsh_1,Marsh_2}. The contour levels are from $9\times 10^{21}$~\pcms\ in steps of $5.25\times 10^{21}$~\pcms. The plus sign indicates the position of the IRAS source. The telescope beam is indicated in the lower left corner.
    }
\label{fig:l1287-maps}
\end{figure}

Several sources have been mapped in these observations. Examples of the maps of L1287 (RNO1) in the lines of some deuterated molecules obtained with the 30-m telescope are presented in Fig.~\ref{fig:l1287-maps}. These maps show significant differences between distributions of the targeted molecules. In general, the DCN peaks are observed towards the temperature peaks, which coincide with locations of luminous IR sources, while other deuterated molecules trace colder regions. This behavior is consistent with the temperature dependencies mentioned above. 


\subsection{The H$^{13}$CN/HN$^{13}$C ratio as a temperature indicator}
\begin{figure}
\begin{minipage}{0.49\textwidth}
    \centering
    \includegraphics[width=\textwidth]{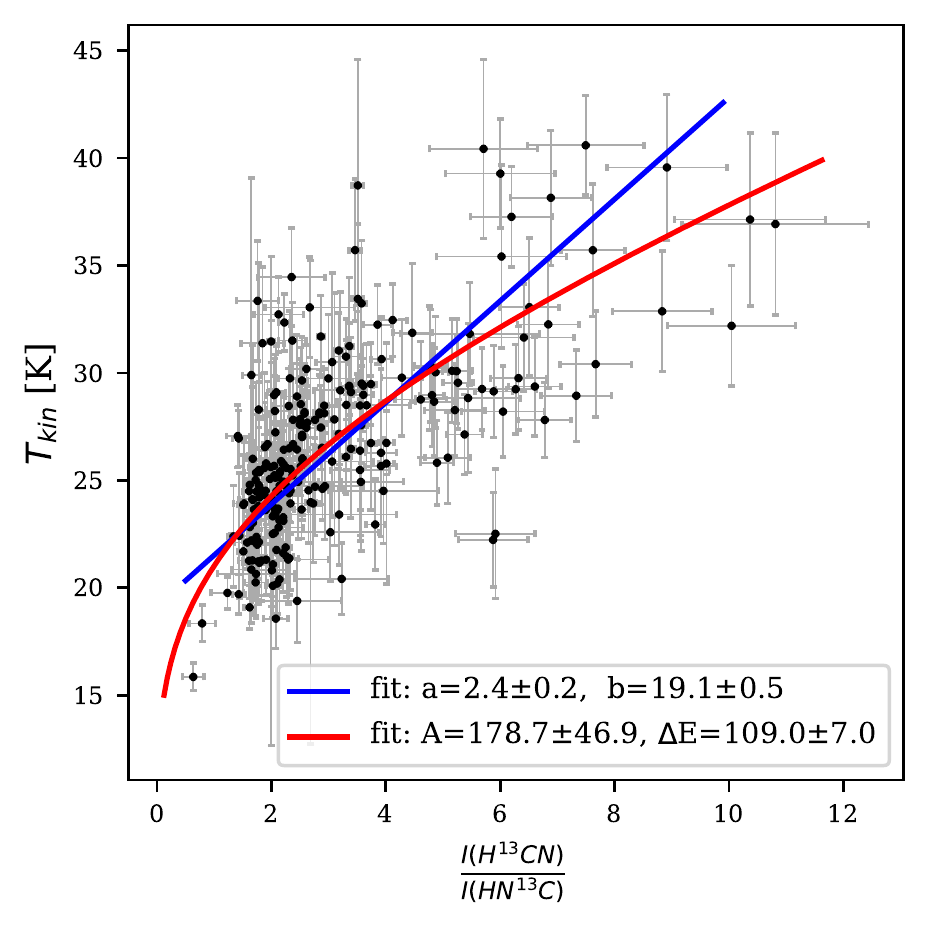}
\end{minipage}
\hfill
\begin{minipage}{0.49\textwidth}
    \centering
    \includegraphics[width=\textwidth]{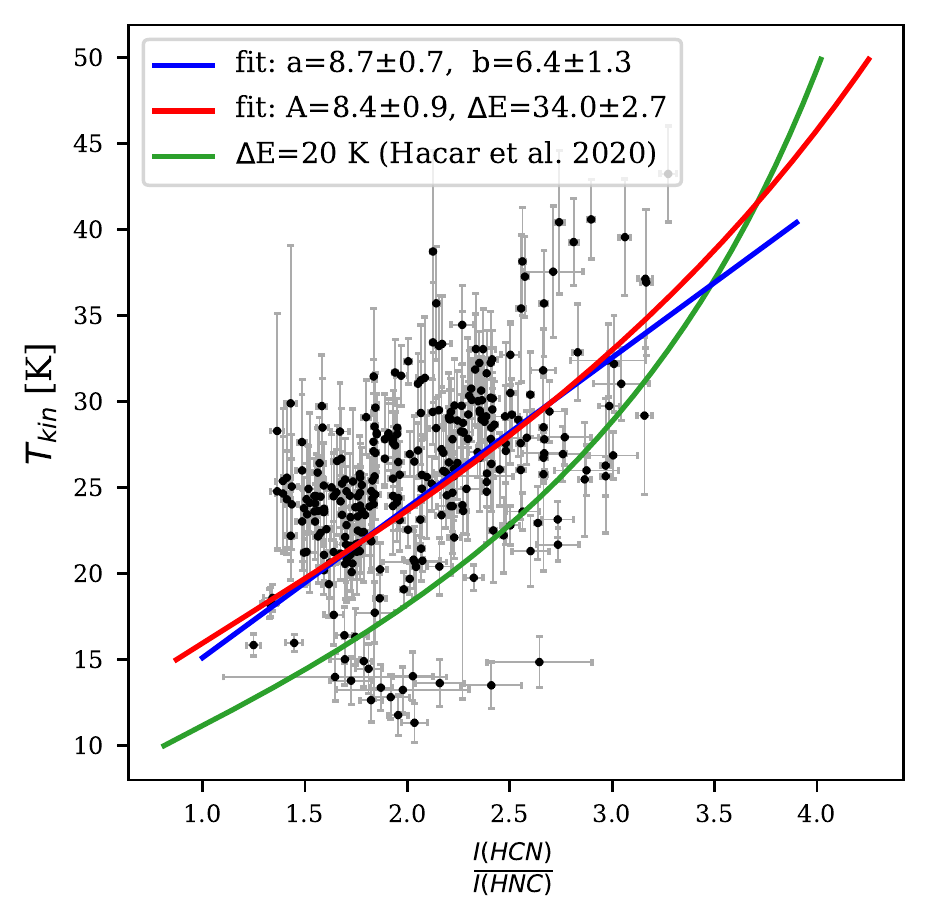}
\end{minipage}
    \caption{Left panel: dependence of the ratio of the integrated intensities of the H$^{13}$CN and HN$^{13}$C lines on the kinetic temperature derived from the CH$_3$CCH and NH$_3$ data. The blue line corresponds to the linear fit $ax+b$, the red curve is $A\times \exp\left( {-\Delta E/T_{kin}} \right)$. The fitting results are shown in the graph. Right panel: same for HCN and HNC. The green curve corresponds to the results presented in \cite{Hacar20} (Pazukhin et al., in preparation).}
    \label{fig:ratio}
\end{figure}

The CH$_3$CCH and NH$_3$ lines observed in our sources are ``traditional'' indicators of the gas kinetic temperature. However, the NH$_3$ maps have been obtained for very few sources only, and the CH$_3$CCH lines are too weak in most sources, at least outside the emission peaks. Recently, the HCN/HNC $J=1-0$ intensity ratio was suggested as a temperature indicator \cite{Hacar20}. As shown in this work, variations of this ratio are driven by the effective destruction and isomerization mechanisms of HNC under low-energy barriers. However, the optical depth in these lines is usually quite high, which may influence this ratio. We analyze the ratio of the presumably optically thin emission in the H$^{13}$CN and HN$^{13}$C $J=1-0$ lines (Pazukhin et al., in preparation). The results are presented in Fig.~\ref{fig:ratio}. They show a good correlation between the H$^{13}$CN/HN$^{13}$C intensity ratio and the gas kinetic temperature. The energy barrier for the HN$^{13}$C destruction is $\sim$100~K, much higher than that derived in \cite{Hacar20} for HNC in the temperature range $\lesssim 40$~K. The results show that the H$^{13}$CN/HN$^{13}$C intensity ratio may be a good indicator of the gas kinetic temperature and we employ this correlation in the further analysis. For the HCN/HNC ratio our results are close to those presented in \cite{Hacar20}, although the derived energy barrier for HNC destruction is somewhat higher (34~K here versus 20~K in \cite{Hacar20}). It is worth noting that we found a very poor correlation between the gas and dust temperatures in our sources. This can be related to insufficient gas density.

\subsection{Physical parameters of the sources and deuteration degrees}

Using temperature estimates as outlined above, we derived the gas volume density and molecular column densities from the $J=1-0$ and $J=2-1$ transitions of DCN, DNC and DCO$^+$ by non-LTE modeling with RADEX. Column densities of their hydrogenated isotopologues were derived from observations of the $J=1-0$ transitions only using the estimates of the temperature and volume density mentioned above. Then, deuteration degrees for these species were obtained. Maps of these parameters have been constructed. The details of these procedures and results will be presented elsewhere (Pazukhin et al., in preparation).

Briefly, gas volume densities in the investigated objects vary in the range $\sim 10^4-10^6$~\pcmc. Typical deuteration degrees are $\sim$10$^{-2}$ at temperatures of 10--20~K., in some cases they reach $\sim$3$\times 10^{-2}$. The deuteration degrees for HCO$^+$ and HNC drop with increasing temperature, while for HCN it is more or less constant. These results are consistent with the findings in the Onsala survey (Sect.~\ref{sec:survey}).

\section{Conclusions}
As a result of a survey of high-mass star forming regions using the 20-m radio telescope of the Onsala Observatory, emission from DCO$^+$ molecules was detected in 16 out of 50, DCN in 17 out of 50, DNC in 15 out of 47, N$_2$D$^+$ in 2 out of 47, and NH$_2$D in 15 out of 50 observed sources.
The spatial distributions of various deuterated molecules are significantly different. DCN emission peaks are observed in the direction of IR sources, while DCO$^+$ and DNC trace colder regions.
The content of deuterated molecules with respect to the main isotopologues (deuteration degree) is $\sim$10$^{-2}$ at temperatures of 10--20~K. The deuteration degrees for HCO$^+$ and HNC drop with increasing temperature, while for HCN and NH$_3$ they are more or less constant at temperatures up to $\sim$50~K. This is consistent with the available chemical models, which show a high efficiency of DCN formation at temperatures up to $\sim$80 K but contradicts the available models for NH$_3$.
The relative abundance of DCO$^+$ drops sharply with increasing linewidth. There is a correlation between line width and temperature.
The H$^{13}$CN/HN$^{13}$C intensity ratio may be a good indicator of the gas kinetic temperature.
The gas volume densities in the investigated objects vary in the range $\sim 10^4-10^6$~\pcmc.

\begin{acknowledgments}
This research was supported by the Russian Science Foundation grant 22-22-00809. This work is partly based on observations carried out under project number 041-19 with the IRAM 30m telescope and on observations with the 100-m telescope of the MPIfR (Max-Planck-Institut für Radioastronomie) at Effelsberg. IRAM is supported by INSU/CNRS (France), MPG (Germany) and IGN (Spain). We are grateful to the IRAM and MPIfR staff for their support during the observations. We also acknowledge support from Onsala Space Observatory for  the provisioning of its facilities/observational support. The Onsala Space Observatory national research infrastrcuture is funded through Swedish Research Council grant No 2017-00648.

\end{acknowledgments}

%
\bibliographystyle{JHEP}
\bibliography{zinchenko}

\end{document}